\documentclass[reviewcopy]{elsart}
\usepackage[latin1]{inputenc}
\usepackage{amssymb}
\usepackage{amsmath}
\usepackage{relsize}
\usepackage{xspace}
\usepackage{url}

\usepackage{epsfig}
\usepackage{graphicx}
\usepackage{subfigure}

\usepackage{array}
\usepackage{multicol}

\usepackage[english]{babel}

\graphicspath{{Fig/}}
\newcommand{\beq}{\begin{equation}}
\newcommand{\eeq}{\end{equation}}
\newcommand{\ben}{\begin{enumerate}}
\newcommand{\een}{\end{enumerate}}
\newcommand{\nid}{\noindent}

\newtheorem{theorem}{Theorem}

\newcommand{\bth}{\begin{theorem}}
\newcommand{\enth}{\end{theorem}}
\newtheorem{assumption}{Assumption}
\newcommand{\bas}{\begin{assumption}}
\newcommand{\eas}{\end{assumption}}

\newcommand\x{\textbf{x}}
\renewcommand\u{\textbf{u}}
\renewcommand\v{{\bf v}}

\newcommand\uiT{u_i^{(T)}}
\newcommand\F{\textbf{F}}

\newcommand\DFx{D\F_\x}
\newcommand\DFTx{D\F^{(T)}_\x}

\newcommand\cW{{\cal W}}
\newcommand\WTp{\cW^{(T+1)}}
\newcommand\WT{\cW^{(T)}}

\newcommand{\XI}{\mbox{\boldmath $\xi$}}
\newcommand\lmT{L_{1}^{(T)}}

\newcommand\siT{s_i^{(T)}}
\newcommand\suT{s_1^{(T)}}

\newcommand\Hinf{\textbf{H}^{(\infty)}}

\newcommand\uaT{\u^{\ast(T)}}

\newcommand\lb{\left\langle}
\newcommand\rb{\right\rangle}
\newcommand\rbT{\rb^{(T)}}

\newcommand\moyxinf{\lb\x\rb^{(\infty)}}

\newcommand\moyuinf{\lb\u\rb^{(\infty)}}

\newcommand\dtl{\Delta^{(T)}[\Lambda]}
\newcommand\bbbr{{\sf I\!R}}
\begin{document}

\begin{frontmatter}
      \title{Effects of Hebbian learning on the dynamics and structure of random networks with inhibitory and excitatory neurons}
      \author[Alchemy]{Beno\^{\i}t Siri}
      \author[ETIS]{,~Mathias Quoy}
      \author[ANIM]{,~Bruno Delord}
      \author[INLN,Odyssee]{,~Bruno Cessac}
      \author[Alchemy]{,~Hugues Berry\corauthref{cor}}
      \corauth[cor]{Corresponding author: hugues.berry@inria.fr}
      \address[Alchemy]{INRIA, Project-Team Alchemy, 4 rue J Monod, 91893 Orsay Cedex, France}
      \address[ETIS]{ETIS, UMR 8051 CNRS-Universit\'e de Cergy-Pontoise-ENSEA, 6 avenue du Ponceau, BP 44, 95014 Cergy-Pontoise Cedex, France}
      \address[ANIM]{ANIM, U742 INSERM - Universit\'e P.M. Curie, 9 quai Saint-Bernard, 75005 Paris, France}
      \address[INLN]{INLN, UMR 6618 CNRS-Universit\'e de Nice, 1361 route des Lucioles, 06560 Valbonne, France}
       \address[Odyssee]{INRIA, Project-Team Odyssee, 2004 Route des Lucioles, 06902 Sophia Antipolis, France}

      \begin{abstract}
The aim of the present paper is to study the effects of Hebbian learning in random recurrent neural networks with biological connectivity, i.e. sparse connections and separate populations of excitatory and inhibitory neurons. We furthermore consider that the neuron dynamics may occur at a (shorter) time scale than synaptic plasticity and consider the possibility of learning rules with passive forgetting. We show that the application of such Hebbian learning leads to drastic changes in the network dynamics and structure. In particular, the learning rule contracts the norm of the weight matrix and yields a rapid decay of the dynamics complexity and entropy. In other words, the network is rewired by Hebbian learning into a new synaptic structure that emerges with learning on the basis of the correlations that progressively build up between neurons. We also observe that, within this emerging structure, the strongest synapses organize as a small-world network. The second effect of the decay of the weight matrix spectral radius consists in a rapid contraction of the spectral radius of the Jacobian matrix. This drives the system through the ``edge of chaos'' where sensitivity to the input pattern is maximal. Taken together, this scenario is remarkably predicted by theoretical arguments derived from dynamical systems and graph theory.
      \end{abstract}
\end{frontmatter}

\section{Introduction}
Neural networks show amazing abilities for information storage and processing, and stimulus-dependent activity shaping. These capabilities are mainly conditioned by the mutual coupling relationships between network structure and neuron dynamics. Actually, learning in neural networks implies that activity guides the way synapses evolve; but the resulting connectivity structure in turn can raise new dynamical regimes. This interaction becomes even more complex if the considered basic architecture is not feed-forward but includes recurrent synaptic links, like in cortical structures. Understanding this mutual coupling between dynamics and topology and its effects on the computations made by the network is a key problem in computational neuroscience, that could benefit from new approaches.\\
In the related field of dynamical systems interacting via complex coupling networks, a large amount of work has recently focused on the influence of network topology on global dynamics (for a review, see~\cite{Boccaletti06}). In particular, much effort has been devoted to understanding the relationships between node synchronization and the classical statistical quantifiers of complex networks (degree distribution, average clustering index, mean shortest path, modularity...)~\cite{Grinstein05,Nishikawa03,Lago00}. The main idea was that the impact of network topology on the global dynamics might be prominent, so that these structural statistics may be good indicators of the global dynamics. This assumption proved however largely wrong so that some of the related studies yielded contradictory results~\cite{Nishikawa03,Hong02}. Actually, synchronization properties cannot be systematically deduced from topology statistics but may be inferred from the spectrum of the network~\cite{Atay06}. Accordingly, many studies have considered diffusive coupling of the nodes~\cite{Hasegawa05}. In this case, the adjacency matrix has real nonnegative eigenvalues, and global properties, such as stability of the synchronized states~\cite{Barahona02}, can easily be inferred from its spectral properties.\\
\indent In this perspective, neural networks can be considered as mere examples of these complex systems, with the particularity that the dynamics of the network nodes (neurons) depends on the network links (synaptic weights), that themselves vary over time as a function of the node dynamics. Unfortunately, the coupling between neurons (synaptic weight) is rarely diffusive, so that the corresponding matrix is not symmetric and may contain positive and negative elements. Hence the mutual coupling between neuron dynamics and network structure remains largely to be understood.\\
Our general objective is to shed light on these interactions in the specific case of random recurrent neural networks (RRNNs). These network models display a rich variety of dynamical behaviors, including fixed points, limit cycle oscillations, quasiperiodicity and deterministic chaos~\cite{Doyon_IJBC93}, that are suspected to be similar to activity patterns observed in the olfactory bulb~\cite{Skarda87,Freeman87}. It is known that the application of biologically-plausible local learning rules (Hebbian rules) reduces the dynamics of chaotic RRNNs to simpler attractors that are specific of the learned input pattern~\cite{Dauce_NN98}. This phenomenon endows RRNNs with associative memory properties, but remains poorly understood.\\
Our previous work showed that the evolution of the network structure during learning can be tracked in numerical simulations via the classical topological statistics from ``complex networks approaches''~\cite{Berry_AB06}. More recently, we devised a mathematical framework for the effects of Hebbian learning on the dynamics, topology and some functional aspects of RRNNs~\cite{Siri07}. This theoretical approach was shown to explain the effect of learning in a ``canonical'' RRNN, i.e. a completely connected network where a neuron projects both excitatory and inhibitory synapses. However, this network type remains rather poorly realistic from a biological point of view.\\
The aim of the present paper is thus to study the effects of a more biological connectivity. In particular, we segregate the neurons into two distinct populations, namely excitatory (projecting only excitatory synapses) and inhibitory (projecting only inhibitory synapses) neurons. Furthermore, the network is sparsely connected and the overall connectivity parameters are fixed to emulate local circuitry in the cortex. We show that the application of Hebbian learning leads to drastic changes in the network dynamics and structure. We also demonstrate that the mathematical arguments mentioned above remain a very useful unifying framework to understand the effects of learning in this system.\\
\section{The model}\label{model}
\subsection{Connectivity}
We consider networks with a total of $N=500$ neurons and random connectivity. Each neuron is either inhibitory (with probability $p_I$) or excitatory (with probability $p_E=1-p_I$) and projects to $p_c N$ randomly chosen postsynaptic neurons (independently of their excitatory or inhibitory nature). Probabilities are taken uniform on $[0,1]$ for the network connectivity $p_c$ and fraction of inhibitory neurons $p_I$. In the present study, we fixed $p_I$ and $p_c$ so as to account for the neural circuitry of a typical neocortical column. Hence, we used $p_I=0.25$~\cite{Kendell00} and $p_c=0.15$~\cite{Markram97,Kalisman05}.
\\The initial weight of each synapse between a postsynaptic neuron $i$ and a presynaptic neuron $j$, $W_{ij}^{(1)}$, is drawn at random, according to a Gamma distribution, whose parameters depend on the nature of the presynaptic neuron $j$. If $j$ is inhibitory, $W_{ij}^{(1)} \sim \mathrm{Gamma}(-\mu_w/n_i,\sigma_w/n_i)$, where $\mathrm{Gamma}(m,s)$ denotes the Gamma distribution with mean $m$ and standard deviation $s$, and $n_i=p_Ip_cN$. If $j$ is excitatory, then $W_{ij}^{(1)} \sim \mathrm{Gamma}(\mu_w/n_e,\sigma_w/n_e)$ where $n_e=p_Ep_cN$. Using Gamma distributions (instead of Gaussian ones, for instance) allows to ensure that inhibitory (excitatory) neurons project only negative (positive) synapses, whatever the values of $\mu_w$ and $\sigma_w$. Thanks to the normalization terms ($n_e$ and $n_i$), the total excitation received by a postsynaptic neuron is \textit{on average} equal to the total inhibition it receives. Hence, in their initial setups (i.e. before learning) our networks are guarantied to conserve the excitation/inhibition balance (on average).
\subsection{Dynamics}
We consider firing-rate neurons with discrete-time dynamics and take into account that learning may occur on a different (slower) time scale than neuron dynamics. Synaptic weights are thus kept constant for $\tau \geq 1$ consecutive dynamics steps, which defines a ``learning epoch''. The weights are then updated and a new learning epoch begins. We denote by $t \geq 0$ the update index of neuron states (neuron dynamics) inside a learning epoch,
while $T \geq 1$ indicates the update index of synaptic weights (learning dynamics).
\\Let $x_i^{(T)}(t) \in [0,1]$ be the mean firing rate of neuron $i$, at time $t$ within the learning epoch $T$. Let $\cW^{(T)}$ be the matrix of synaptic weights at the $T$-th learning epoch and $\XI$ the vector $\left(\xi_i\right)_{i=1}^N$. Then the discrete time neuron dynamics (\ref{eq:dyn}) writes:
\begin{equation}\label{eq:dyn}
    x_i^{(T)}(t+1) = f\left(\sum_{j=1}^N W_{ij}^{(T)} x_j^{(T)}(t)+ \xi_i \right).
\end{equation}
Here, $f$ is a sigmoidal transfer function ($f(x)=1/2\left(1+\tanh(gx)\right)$).
The output gain $g$ tunes the nonlinearity of the function and mimics the excitability of
the neuron. $\xi_i$ is a (time constant) external input applied to neuron $i$ and the vector $\XI$ is the ``pattern'' to be learned. $W_{ij}^{(T)}$ represents the weight of the synapse from presynaptic neuron $j$ to postsynaptic neuron $i$ during learning epoch $T$. Finally, at the end of one learning epoch, the neuron dynamics indices are reset: $x_i^{(T+1)}(0)=x_i^{(T)}(\tau), \forall i$.
\subsection{Learning}
In the present work, we used the following Hebbian learning rule (see~\cite{Siri07} for justifications):
\begin{equation}\label{eq:LRule}
W_{ij}^{(T+1)}=\lambda W_{ij}^{(T)}+s_j \frac{\alpha}{N}m_i^{(T)}m_j^{(T)}H\left( m_j^{(T)}\right).
\end{equation}
where $\alpha$ is the learning rate, $s_j=+1$ if $j$ is excitatory and $-1$ if it is inhibitory and $H$ denotes the Heaviside step function ($H(x)=0$ if $x<0$, $1$ else). The first term in the right-hand side (RHS) member accounts for passive forgetting, i.e.
 $\lambda \in [0,1]$ is the forgetting rate. If $\lambda < 1$ and $m_i$ or $m_j=0$ (i.e. the pre- or postsynaptic neurons are silent, see below), eq.(\ref{eq:LRule}) leads to an exponential decay of the synaptic weights (hence passive forgetting). Another important consequence of this rule choice is that if $\lambda<1$, the weights are expected to converge to stationary values. Hence $\lambda<1$ also allows avoiding divergence of the synaptic weights. Note that there is no forgetting when $\lambda=1$.\\
The second term in the RHS member of eq.(\ref{eq:LRule}) generically accounts for activity-dependent plasticity, i.e. the effects of the pre- and postsynaptic neuron firing rates. In our model, this term depends on the \textit{history} of activities through the time-average of the firing-rate:
\begin{equation}\label{eq:mi}
m_i^{(T)}=\frac{1}{\tau}\sum_{t=1}^\tau x_i^{(T)}(t) -d_i,
\end{equation}
where $d_i \in [0,1]$ is a threshold that we set to $d_i=0.10, \, \forall i$ in the present study. A neuron $i$ will thus be considered active during learning epoch $T$ whenever $m_i^{(T)}>0$ (i.e. whenever its average firing rate has been $>10 \%$ of the maximal value), and silent else.\\
Note that definition (\ref{eq:mi}) actually encompasses several cases. If $\tau=1$, weight changes depend only on the instantaneous firing rates, while if $\tau \gg 1$, weight changes depend on the mean value of the firing rate, averaged over a time window of duration $\tau$ in the learning epoch. In many aspects the former case can be considered as genuine plasticity, while the latter may be related to meta-plasticity~\cite{Metaplasticity}. In this paper, we used $\tau = 10^4$. Finally, weights cannot change their sign. Note however that this setup does not have a significant impact on the present results.
\section{Results}
\subsection{Spontaneous dynamics}
\begin{figure}[h!]
\centerline{\includegraphics[scale=0.90]{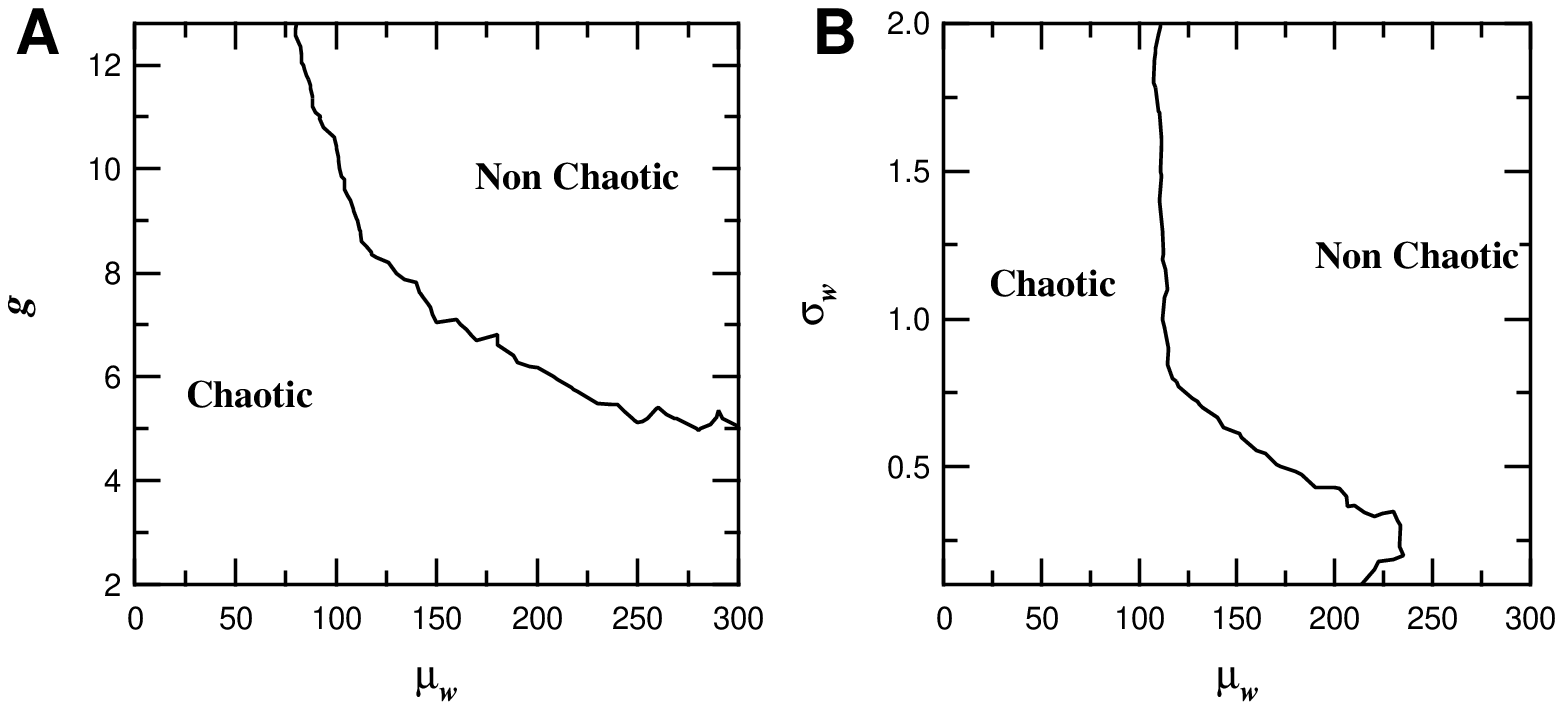}}
\caption{Phase diagram for the spontaneous dynamics eq.(\ref{eq:dyn}). The full line represents the boundary between chaotic and non chaotic dynamics (i.e. the isocurve $L_1=0$ where $L_1$ is the largest Lyapunov exponent). Shown are projection in (\textit{A}) the  $(g,\mu_w)$ plan with $\sigma_w=1.0$ or (\textit{B}) the $(\sigma_w,\mu_w)$ parameter plan with $g=10.0$. Other parameters were: $\xi_i=0.0 \, \forall i=1 \ldots N$, $p_c=0.15, p_I=0.25$ and $N=500$.}
  \label{fPhaseDiag}
\end{figure}

We first present simulation results on the spontaneous dynamics of the system, i.e. the dynamics eq.(\ref{eq:dyn}) \textit{in the absence of learning}. The phase diagrams in fig.~\ref{fPhaseDiag} locate regions in the parameter space for which chaotic dynamics are observed in simulations. Each panel shows the isocurve $L_1=0$ (where $L_1$ is the largest Lyapunov exponent) that represents the boundary between chaotic ($L_1>0$) and non chaotic ($L_1<0$) dynamics.\\
It is clear from this figure that chaotic dynamics are found for large parts of the parameter space. Generally speaking, chaotic behaviors disappear when the average weight $\mu_w$ increases, which may be related to an increase of the average neuron saturation. A more surprising observation is that chaotic dynamics tends to \textit{disappear} when the gain of the transfer function $g$ is increased. This behavior is in opposition to the behavior observed with classical random recurrent networks with homogeneous neurons (where each neuron has both excitatory and inhibitory projections). In the latter models (and even in related two-populations models, see~\cite{Dauce_BC2002}), chaotic dynamics usually appear for increasing values of $g$ (see e.g.~\cite{CS}).\\
This is a very interesting property of the spontaneous dynamics in our model, whose understanding is however out of the scope of the present paper and is left for future work. In the framework of the present study, these phase diagrams mainly allow locating suitable parameters for the initial conditions of our networks. We wish the initial dynamics to provide a wide range of possible dynamical regimes, a large (KS) entropy and self-sustaining dynamics. For these reasons, we set our initial dynamics inside the chaotic region, and fix $\mu_w=50, \, \sigma_w=1.0, \, g=10$ and $N=500$. The initial weight distribution will thus consist in a Gamma distribution with effective average $-2.67$ and s.d. $0.053$ for inhibitory synapses, and  $0.89$ and $0.018$, respectively, for excitatory ones.
\subsection{Structure modifications}\label{structure}
In this section, we study the changes in the network structure induced by the learning rule described by eq.(\ref{eq:LRule}).
\subsubsection{Adjacency matrix}\label{A}
\begin{figure}[h!]
\centerline{\includegraphics[width=\textwidth]{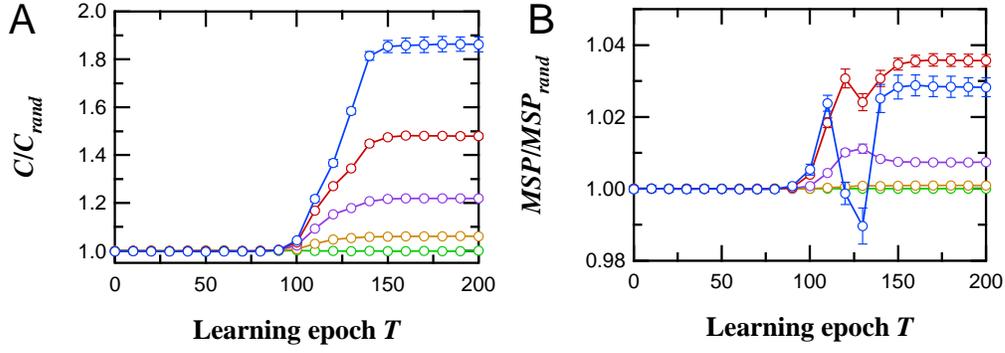}}
\caption{Evolution of the normalized structural statistics during learning with rule eq.(\ref{eq:LRule}). Values are averages over $20$ different realizations of the network (random initial firing rates and synaptic weights). The values of the threshold $\theta$ are, from bottom to top in each panel, 100\%, 87\%, 73\%, 60\% and 47\%. (\textit{A}) Normalized clustering index $C^{(T)}(\theta)/C^{(T)}_{rand}(\theta)$. (\textit{B}) Normalized mean-shortest path $MSP^{(T)}(\theta)/MSP^{(T)}_{rand}(\theta)$. Bars are $\pm 1$ standard deviation. Other parameters were: $\lambda=0.90$, $\alpha=5 \times 10^{-3}, \, g=10$, $\xi_i=0.010 \sin\left( 2 \pi i/N \right) \cos\left( 8 \pi i/N \right) \, \forall i=1 \ldots N$, $\mu_w=50, \, \sigma_w=1.0$ and $N=500$.}
  \label{fstruct}
\end{figure}

The adjacency matrix of a graph gives information about its connectivity and can be extracted from the weight matrix $\cW$ by thresholding and binarization. We applied a simple relative thresholding method that consists in keeping only the absolute values of the $\theta$ percent highest weights (again, in absolute value) from the nonzero connections in $\cal W$. Hence gradual decrease of $\theta$ enables to progressively isolate the adjacency network formed by the most active weights only. The resulting matrix is then binarized and symmetrized, yielding the adjacency matrix $\cal{A}(\theta)$ whose elements $a_{ij}(\theta)$ indicate whether $i$ and $j$ are connected by a synapse with a large ($> \theta$) weight (either inhibitory or excitatory), compared to the rest of the network. We limit the range of $\theta$ values to ensure that not more that $10 \%$ of the neurons get disconnected from the network by the thresholding process.\\
To characterize the topology of these matrices, we computed the two main quantifiers used in ``complex networks'' approaches, namely the clustering index and the mean shortest path (see~\cite{Siri07} for formal definitions). The clustering index $C$ is a statistical quantifier of the network structure and reflects the degree of ``cliquishness'' or local clustering in the network~\cite{Watts98}. It expresses the probability that two nodes connected to a third one are also connected together and thus can be interpreted as the density of triangular subgraphs in the network. The mean shortest path ($MSP$) is the average, over all nonidentical neurons pairs $(i,j)$, of the smallest number of synapses one must cross to reach $i$ from $j$.\\
These two quantifiers are usually informative only when compared to similar measures obtained from reference random networks~\cite{Watts98}. Here, to build reference networks, we start with the weight matrix at learning epoch $T$, ${\cal W}^{(T)}$ and rewire it at random but preserving the inhibitory/excitatory nature of the neurons. Hence for each element $W_{ij}^{(T)}$, we choose (uniformly) at random another element $W_{kl}^{(T)}$ \textit{with the same sign}, and exchange their values. We then compute the clustering index and mean shortest path of the resulting rewired network, and average the obtained values over $15$ realizations of the rewiring process, yielding the reference values $C_{rand}(\theta)$ and $MSP_{rand}(\theta)$.\\
Figure~\ref{fstruct}A \& B show simulation results for the evolution of $C^{(T)}(\theta)/C^{(T)}_{rand}(\theta)$ and $MSP^{(T)}(\theta)/MSP^{(T)}_{rand}(\theta)$ during learning. The distribution of the initial weights over the network being random, the resulting adjacency matrix ${\cal A}^{(1)}(\theta)$ is essentially a random network, i.e. one expects $C^{(1)}(\theta)/C^{(1)}_{rand}(\theta) \approx 1$ and $MSP^{(1)}(\theta)/MSP^{(1)}_{rand}(\theta) \approx 1, \, \forall \theta$. This is confirmed in fig.~\ref{fstruct}: during the approximately first $100$ learning epochs, both network statistics remain around $1$.\\
The situation however changes for longer learning epochs. For $T \gtrsim 100$, the relative MSP remains essentially $1$ for all thresholds $\theta$ (less than $4\%$ variation, fig.~\ref{fstruct}B). Hence, the average minimal number of synapses linking any two neurons in the network remains low, even when only large synapses are considered. Conversely, the clustering index (fig.~\ref{fstruct}A) increases at $T>100$ for the stronger synapses and reaches a stable value that is up to almost two twofold the value found in the reference random networks. Hence, if one considers the strong synapses at long learning epochs, the probability that the neighbors of a given neuron are themselves interconnected is almost twofold higher than if these strong synapses were laid at random over the network. In other terms, the learning rule yields correlations among the largest synapses at long learning epochs.
\\In the literature related to ``complex networks'', networks with a larger clustering index but a similar MSP with respect to a comparable reference random network, are referred to as \textit{small-world} networks. Hence, the learning rule eq.(\ref{eq:LRule}) organizes strong synapses as a ``small-world'' network.\\
Emerging experimental evidence shows that numerous brain anatomical and functional connectivity networks at several length scales indeed display a common small-world connectivity (for a recent review, see~\cite{Bassett06}). These include quantifications of the physical~\cite{Shefi04} or functional~\cite{Bettencourt07} connectivity of neuronal networks grown in vitro; quantifications of the anatomical connectivity of \textit{Caenorhabditis elegans} full neural system~\cite{Watts98} or, at larger scale, cortico-cortical area connectivity maps in macaque, cat~\cite{SpornsZwi04} and more recently human~\cite{He07}; and quantitative studies of \emph{functional} human brain connectivity based on MEG~\cite{Stam04}, EEG~\cite{Michel06} or fMRI data~\cite{Achard06,Eguiluz05}. Current hypothesis for the frequent observation of small-world connectivity in real biological networks state that they may result from natural evolution (small-world networks are generally thought to minimize wiring length while preserving low energy costs and high local integration of the neurons~\cite{Bassett06,Kaiser06}).\\
An alternative hypothesis could be that small-world networks are emerging properties of neural networks subject to Hebbian learning. In favor of this possibility, small-world connectivity has recently been shown to arise spontaneously from spiking neuron networks with STDP plasticity and total connectivity~\cite{Shin05} or with correlation-based rewiring~\cite{Kwok07}. Hence our present findings tend to strengthen this hypothesis.\\
However, these kinds of interpretation should be taken with great care. For instance, it is easy to find learning rules similar to eq.(\ref{eq:LRule}) or areas in the parameter space, for which the network, even at long learning times, only slightly deviates from its initial random organization (see e.g.~\cite{Siri07}). Hence, emergence of small-world connectivity, even in computational models of neural networks (i.e. not to speak about real neural networks), may be restricted to certain areas of the learning rule parameter space.\\
More importantly, these indicators in fact give no obvious clue about the mutual coupling between global dynamics and the network structure. Hence, in our case at least, the classical statistics of the ``complex networks'' do not provide causal explanation for the dynamical effects of learning. For instance, it does not help understand why dynamics complexity systematically decreases during learning. However, the adjacency matrix is not the only viewpoint from which the network structure can be observed (see~\cite{Siri07} for a discussion). In the following, we examine what information can be obtained if the structure is observed at the level of the Jacobian matrices.
\subsubsection{Jacobian matrices.}\label{DF}
\begin{figure}[h!]
  \centerline{\includegraphics[width=1.2\textwidth]{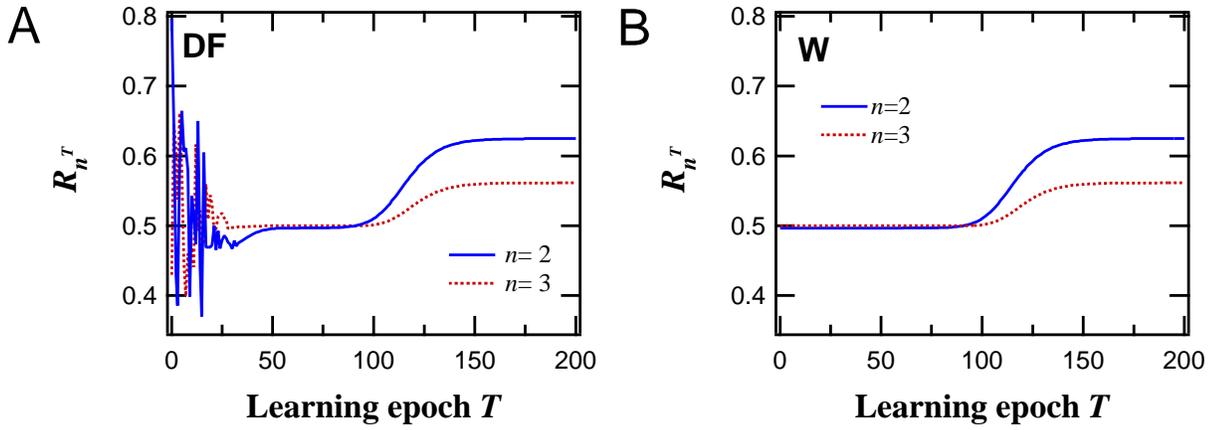}}
\caption{Evolution of the weighted-fraction of positive feedback loops $R_n^{(T)}$ for loops in {$D\F$} (A) and {$\cW$} (B) and circuit length $n=2$ (thick line) and $n=3$ (dotted line). Values are averages over 20 different networks using $\lambda=0.90$. Standard deviations are omitted for readability purpose. See text for definitions. All other parameters as in fig.~\ref{fstruct}.}
  \label{floops}
\end{figure}

Denote by $\F$ the function $\F: \bbbr^N \to \bbbr^N$ such
that $F_i(\x)=f(x_i)$. In our case, the components of the \textit{Jacobian matrix} of $\F$ at $\x$, denoted by $\DFx$ are given by:
\beq\label{DFij}
\frac{\partial F_i}{\partial x_j} =f'(\sum_{k=1}^N W_{ik}x_k+\xi_i)W_{ij}=f'(u_i)W_{ij}.
\eeq
Thus it displays the following specific structure:
\beq\label{DFx}
\DFx=\Lambda(\u)\cW,
\eeq
\nid with:
\beq\label{Lambda}
\Lambda_{ij}(\u)=f'(u_i)\delta_{ij}.
\eeq
Note that $\DFx$ depends on $\x$, contrarily to $\cW$. Generally speaking, $\DFx$ gives the effects of perturbations at the linear order. To each Jacobian matrix $\DFx$ one can associate a graph, called ``the graph of linear influences''. To build this graph, one draws an oriented link $j \rightarrow i \quad \mathrm{iff} \quad \frac{\partial f(u_i)}{\partial x_j} \neq 0$.
The link is positive if $\frac{\partial f(u_i)}{\partial x_j} > 0$ and negative if $\frac{\partial f(u_i)}{\partial x_j} < 0$. A detailed presentation of the properties of the graph of linear influences can be found in~\cite{CS,Siri07}. We just recall here that this graph contains \textit{circuits or feedback loops}. If $e$ is an edge, we denote by $o(e)$ the origin of the edge and $t(e)$ its end. Then a feedback loop is a sequence of edges $e_1, . . . ,e_k$ such that $o(e_{i+1}) = t(e_i)$, $\forall i = 1 . . . k -1$, and $t(e_k) = o(e_1)$. A feedback loop is said positive (negative) if the product of its edges is positive (negative).\\
In general, positive feedback loops are expected to promote fixed-point stability~\cite{Hirsch89} whereas negative loops usually generate oscillations~\cite{Thomas81,Gouze98}. In a model such as eq.(\ref{eq:dyn}) the weight of a loop $k_1,k_2, \dots k_n,k_1$ is given by
$\prod_{l=1}^{n} W_{k_{l+1}k_l}f'(u_{k_l})$, where $k_{n+1}=k_1$. Therefore, the weight of a loop is the product of a ``topological'' contribution ($\prod_{l=1}^{n} W_{k_{l+1}k_l}$) and a dynamical one ($\prod_{l=1}^{n} f'(u_{k_l})$).\\
We measured the evolution of feedback loops during learning via the weighted-fraction of positive circuits in the Jacobian matrix,
 $R_n^{(T)}$, that we defined as
\begin{equation}
\label{eq:powerindex}
R_n^{(T)}=\frac{\sigma^{+(T)}_n}{|\sigma^{+(T)}_n|+|\sigma^{-(T)}_n|}
\end{equation}
where $\sigma^{+(T)}_n$ (resp. $\sigma^{-(T)}_n$) is the sum of the weights of every positive (resp. negative) feedback loops of length $n$ in the Jacobian network at learning epoch $T$. Hence $R_n^{(T)} \in [0,1]$. If its value is $> 0.5$, the positive feedback loops of length $n$ are stronger (in total weight) than the negative ones in the network. We computed the weighted-fraction of positive feedback loops for length $n=2$ and $n=3$ (i.e. $R_2^{(T)}$ and $R_3^{(T)}$).\\
The evolutions of $R_2^{(T)}$ and $R_3^{(T)}$ are presented in fig.~\ref{floops}A. During the first $\approx 20$ learning epochs, the time course of these quantities are highly noisy (and the corresponding standard deviation very large), so that we could not interpret them conclusively. However, a $T \approx 25$ learning epochs, $R_2^{(T)}$ stabilizes to values $<0.5$ ($R_2^{(1)} \approx 0.47$), indicating a slight imbalance in favor of negative feedback loops over positive ones. According to the above theoretical considerations, this indicates a trend toward complex oscillatory dynamics. This viewpoint may be considered another perspective to explain the initial chaotic dynamics. Note however that the initial imbalance in circuits of length-3 is much more modest, $R_3^{(1)} \approx 0.497$.\\
When $25< T <50$, $R_2^{(T)}$ increases and converges to $\approx 0.50$. A dynamical interpretation would be that the corresponding dynamics attractors become progressively less chaotic and more periodic. This is exactly the behavior observed in the simulations (see fig.~\ref{fLyap}B). Hence in spite of the huge fluctuations observed at the beginning, the study of the feedback loops in the Jacobian matrix offers a useful interpretation to the reduction of dynamics induced by learning at short learning epochs.\\
Upon further learning, $R_2^{(T)}$ and $R_3^{(T)}$ remain constant at $0.5$ up to $T \approx 100$ learning epochs. Thus, these quantities do not detect variations in the balance between positive and negative feedback loops for $50<T<100$. However, at longer times ($T>100$), $R_2^{(T)}$ and $R_3^{(T)}$ both increase abruptly and rapidly reach $\approx 0.62$ for $R_2^{(T)}$ and $\approx 0.56$ for $R_3^{(T)}$. Hence, at long learning epochs, the system switches to a state where positive feedback loops hold a significantly larger weight as compared to negative ones. Note that the time course of these indicators for $T>25$ closely follows the time course of the relative clustering index (fig.~\ref{fstruct}A). The causal relation between these two phenomena is however not obvious.\\
Because of the particular form of the Jacobian matrix in our system, the sign of a feedback loop is given by the sign of the weights along it (see above). We thus proceeded (fig.~\ref{floops}B) to the computation of the evolution of the weighted-fraction for feedback loops computed in $\cal{W}$, i.e. we compute here the weight of a feedback loop $e_1, \ldots ,e_k$ as the product of the \textit{synaptic} weights of its edges, thus independently of the neuron state. The evolution of the weighted-fraction of positive feedback loops in $\cal{W}$ does not account for the initial imbalance observed in the feedback loops of $D\F$. However, its evolution at long times is remarkably identical to that measured in $D\F$. Thus, the weighted-fraction of positive feedback loops in $\cal{W}$ is able to account for at least part of the evolution of the dynamics and represents a link between purely structural and purely dynamical aspects. However, more information can be extracted by a more dynamical approach.
\subsection{Dynamical perspective}\label{dynamics}
Starting from spontaneous chaotic dynamics, application of the Hebbian learning rule (\ref{eq:LRule}) in our sparse two-populations model leads to dynamics simplification, as in the case of completely-connected, one-population random recurrent neural networks~\cite{Dauce_NN98}. Figure~\ref{fLyap}B shows the network-averaged neuron dynamics obtained at different learning epochs. The dynamics, initially chaotic ($T=1$), gradually settles onto a periodic limit cycle ($T=270$), then on a fixed point attractor at longer learning epochs (see e.g. $T=290$ in this figure). This evolution of the global dynamics is a typical example of the reduction of the attractor complexity due to the mutual coupling between weight evolution and neuron dynamics.\\
In~\cite{Siri07}, we developed a theoretical approach derived from dynamical systems and graph theory and evidenced that it explains this reduction of complexity in homogenous (single population) recurrent neural networks. We shall show thereafter that it also provides a useful framework for the present model. Below, we first summarize the main results obtained from this mathematical analysis (for details, see~\cite{Siri07}).
\subsubsection{Main theoretical results}
The first prediction of our approach is that Hebbian learning rules contract the norm of the weight matrix $\cal W$. Indeed, we could compute the following upper bound:
\begin{equation} \label{eq:ContracW}
\| \WTp \| \leq \lambda^T \|{\cW}^{(1)}  \| + \frac{\alpha}{N}\frac{1}{1-\lambda} C,
\end{equation}
where $\| \|$ is the operator norm (induced e.g. by Euclidean norm) and $C$ a constant depending on the details of the rule. Hence the major effect of the learning rule is expected to be an exponentially fast contraction of the norm (or equivalently the spectral radius) of the weight matrix, which is due to the term $\lambda$, i.e. to passive forgetting ($\lambda<1$).\\
The next prediction concerns the spectral radius of the Jacobian matrix. Indeed, one can derive a bound for the spectral radius of $\DFTx$:
\begin{equation}\label{eq:contractionDF}
|\mu^{(T)}_1(\x)| \leq \max_{i} f'(\uiT) \| \WT\|.
\end{equation}
This equation predicts a contraction of the spectrum of $\DFTx$ that can arise via two effects: either the contraction of the spectrum of
$\WT$ and/or the decay of $\max_{i} f'(u_i)$, which arises from saturation in neuron activity. Indeed,
$f'(u_i)$ is small when $x_i$ is saturated to $0$ or $1$, but large whenever its synaptic inputs are intermediate,
 i.e. fall into the central part of the sigmoid $f(u_i)$. We emphasize that when $\lambda=1$, $\WT$ and $\u^{(T)}$ can diverge and lead $ \max_{i} f'(\uiT)$ to vanish. Hence the spectral radius of the Jacobian matrix can decrease even in the absence of passive forgetting. In all cases, if the initial value of $|\mu^{(T)}_1(\x)|$ is larger than $1$, eq.(\ref{eq:contractionDF}) predicts that the spectral radius may decrease down to a value $<1$. Note that in discrete time dynamical systems the value $|\mu^{(T)}_1(\x)|=1$ locates a \textit{bifurcation} of the dynamical system.\\
According to our present setting, the largest Lyapunov exponent, $\lmT$ depends on the learning epoch $T$. We were able to show that:
\begin{equation}\label{eq:L1decay}
\lmT \leq \log(\| \WT \|) + \left<\log(\max_i f'(u_i)) \right>^{(T)},
\end{equation}
where $\left<\log(\max_i f'(u_i))  \right>^{(T)}$ denotes the time average of $\log(\max_i f'(u_i))$, in the learning epoch $T$ (see~\cite{Siri07} for formal definitions). The second term in the RHS member is related to the saturation of neurons. The first one states that $\lmT$ will decrease if the norm of the weight matrix $\| \WT \|$ decreases during learning, resulting in a possible transition from chaotic to simpler attractors.\\
Let $u_i^{(T)}(t)=\sum_{j=1}^N W_{ij}^{(T)} x_j^{(T)}(t)+ \xi_i$, the local field (or membrane potential) of neuron $i$ at dynamics step $t$ within learning epoch $T$. Our theoretical work also showed that provided $\lambda<1$, the vector $\mathbf{u}=\left( u_i\right)_{i=1}^N$ converges to a fixed point as $T \rightarrow +\infty$:
\begin{equation}\label{eq:Uinf}
\moyuinf =
\XI + \Hinf,
\end{equation}
\nid where:
\beq\label{Hinf}
\Hinf= \frac{\alpha}{N\left( 1-\lambda \right) }\Gamma^{(\infty)} \moyxinf.
\eeq
Therefore, the asymptotic local field is predicted to be the sum of the input pattern plus an additional term $\Hinf$, which accounts for the \textit{history} of the system and can be weak or not, depending on the exact learning rule and system history.\\
In~\cite{Siri07}, we studied the effects of Hebbian learning in a completely connected ($p_c=1$) one-population network (i.e. where each neuron can project inhibitory (negative) and excitatory (positive) synapses) and showed that these analytical arguments explain and describe results of the system simulation with a very good accuracy.\\
While the model studied in the present work is much more compatible with our knowledge of biological neural networks, it is very different from the model studied in~\cite{Siri07}. In the present model, the connectivity is (severely) sparse and the neurons are segregated in two distinct groups, with distinct synaptic properties. Furthermore, the learning rule eq.(\ref{eq:LRule}) is also more complex. Hence, it is not clear whether the above theoretical arguments account for the current case. In particular, these arguments mainly provide upper bounds, whose quality is not guarantied. In the following sections, we present simulation results about the influence of learning on the network dynamics and function, using our theoretical framework as an oracle.\\
\subsubsection{Dynamics evolution in the sparse 2-populations model.}\label{lyapunov}
\begin{figure}
\centering
  \includegraphics[width=\textwidth]{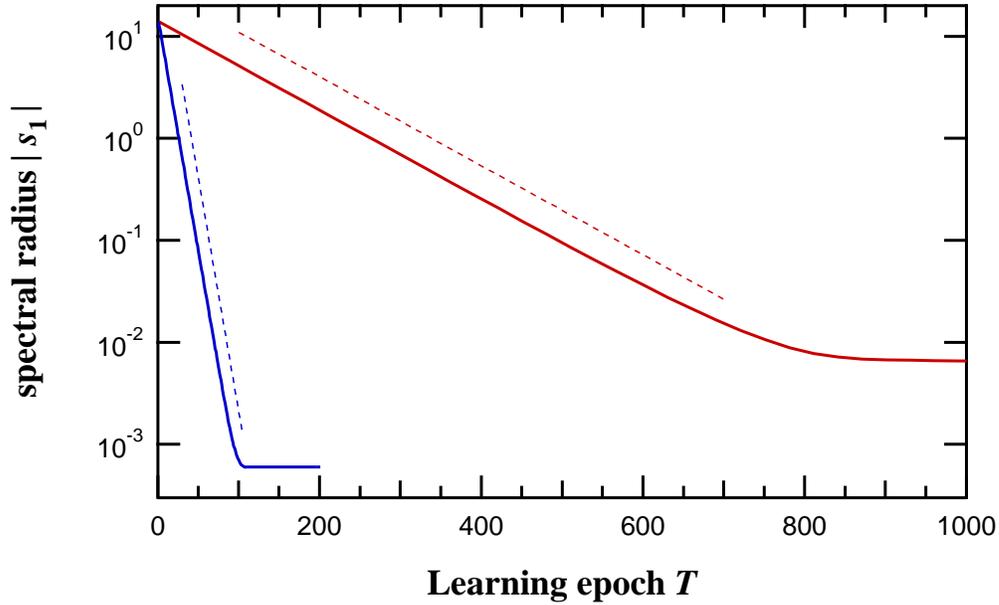}
\caption{Contraction of the spectral radius of ${\cal W}$. The evolution during learning of the norm of ${\cal W}$ largest eigenvalue, $|s_1^{(T)}|$ (thick lines) is plotted on a log-lin scale for $\lambda=0.90$ (bottom) or $0.99$ (top). Each curve is an average over 20 realizations with different initial conditions (initial weights and neuron states). For clarity, standard deviations are omitted but are always $<20\%$ of the average. Dashed thin lines are plots of exponential decreases with equation $g(T)\propto\lambda^T$. All other parameters as in fig.~\ref{fstruct}.}
\label{fSpW}
\end{figure}

\begin{figure}
\centering
  \includegraphics[width=\textwidth]{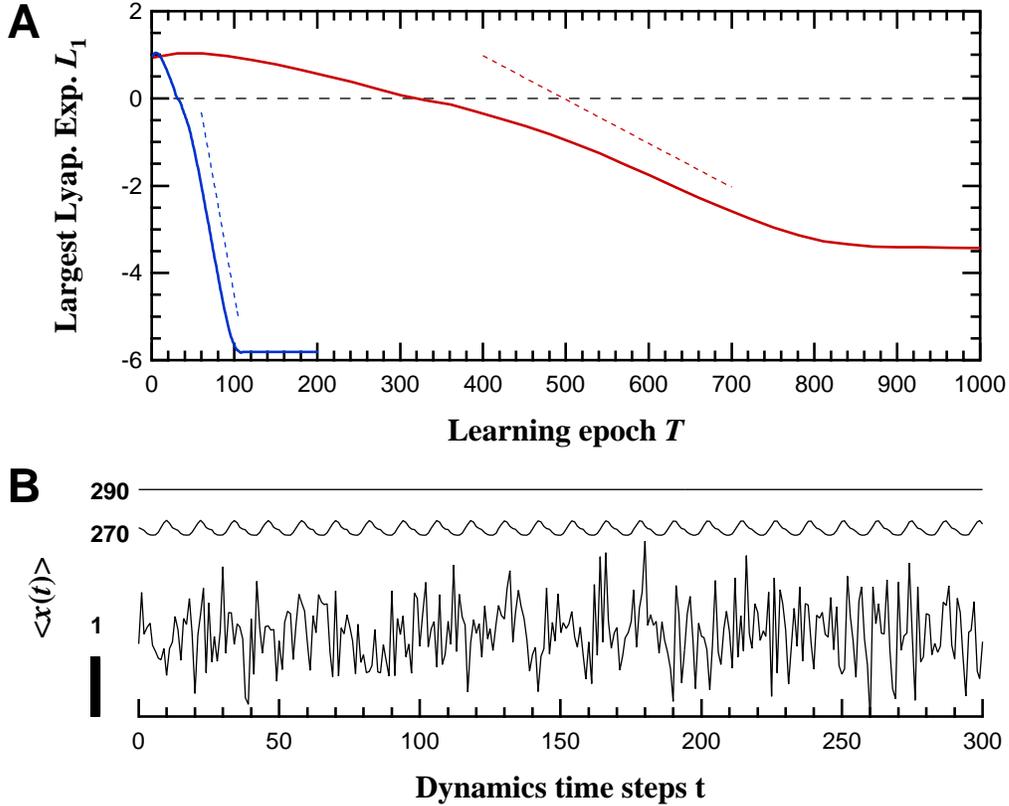}
\caption{Reduction of the dynamics complexity from chaotic to periodic and fixed point. (\emph{A}) Evolution of the largest Lyapunov exponent $L_1$ (full thick lines) for $\lambda=0.90$ (bottom) or $0.99$ (top). Each value is an average over 20 realizations with different initial conditions (initial weights and neuron states). The thin dashed lines illustrate decays as $g(T) \propto T\log(\lambda)$. (\emph{B}) Examples of network dynamics when learning is stopped at (from bottom to top) $T=1$ (initial conditions), $270$ or $290$ and for $\lambda=0.99$. These curves show the network-averaged state $\lb x^{(T)}(t) \rb=1/N \sum_{i=1}^N x_i^{(T)}(t)$ and are shifted along the y-axis for readability. The height of the vertical black bar represents an amplitude of $0.1$. All other parameters are as in fig.~\ref{fstruct}.}
\label{fLyap}
\end{figure}

\begin{figure}
\centering
  \includegraphics[width=\textwidth]{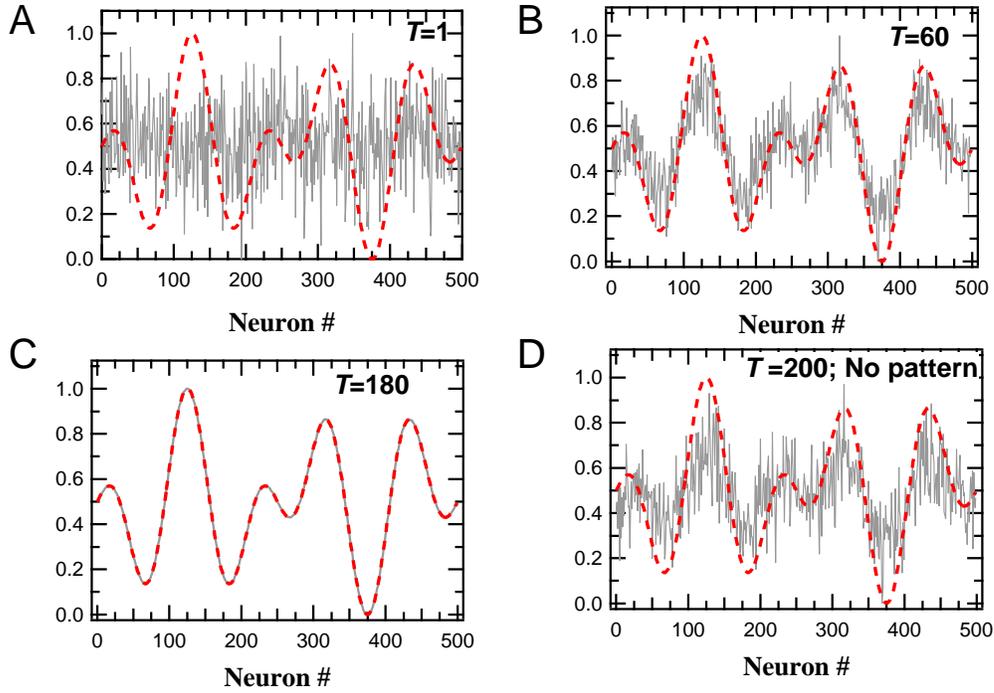}
\caption[long caption]{\label{figu} Alignment of the local field $\u=\cal{W} \x+\XI$ (thin gray line) with the input pattern $\XI$ (thick dashed line). Snapshot are presented at $T=1$ (\textit{A},
 initial conditions), $T=60$ (\textit{B}), $T=180$ (\textit{C}) and $T=200$ with pattern removed (\textit{D}) learning epochs. Each curve plots averages over 20 realizations (standard deviations are omitted for comparison purposes), and every vector has been normalized to $[0, 1 ] $ for clarity. $\lambda=0.90$ and all other parameters as in fig.~\ref{fstruct}}
\end{figure}

Figure~\ref{fSpW} shows the evolution of the spectral radius of ${\cal W}$, $|s_1^{(T)}|$ for $\lambda=0.90$ or $0.99$ in simulations of our sparse two-populations model with dynamics eq.(\ref{eq:dyn}) and learning rule eq.(\ref{eq:LRule}). Let $\siT$ be the eigenvalues of $\WT$, ordered such that $|\suT| \geq |s_2^{(T)}| \geq \dots \geq \siT \geq \dots$. Since $|\suT|$, the spectral radius of $\WT$, is smaller than $\|\WT \|$ one has from eq.(\ref{eq:ContracW}):%
\begin{equation} \label{eq:Contracs1}
|s^{(T+1)}_1| \leq \lambda^T \|{\cW}^{(1)}  \| + \frac{\alpha}{N}\frac{1}{1-\lambda} C.
\end{equation}
It is clear from this figure that in both cases the spectral radius decreases exponentially fast, with a rate that is very close to the prediction of the theory (i.e. $\propto \lambda^T$). Hence, the decay predicted by our analytical approach (eq.\ref{eq:ContracW}) is obviously observed in the simulations. Note that the clear trend in the simulation results for a decay proportional to $\lambda^T$, even tells us that the bound in (\ref{eq:Contracs1}) is indeed very good.\\
Figure~\ref{Fsenspatt} shows (among other curves) the evolution of $|\mu^{(T)}_1(\x)|$ (dashed thin line). This figure confirms that the theoretical prediction about the decay of $|\mu^{(T)}_1(\x)|$ (eq.\ref{eq:contractionDF}) is also valid for this model. Hence, eq.(\ref{eq:contractionDF}) opens up the possibility that learning drives the system through bifurcations. This aspect is studied below (section~\ref{sensitivity}).\\
We now turn to directly study how the attractor complexity changes during learning. This information is provided by the computation of the largest Lyapunov exponent. Note that another canonical measure of dynamical complexity is the Kolmogorov-Sinai (KS) entropy which is bounded from above by the sum of positive Lyapunov exponents. Therefore, if the largest Lyapunov exponent decreases, the KS entropy decreases as well.\\
Figure~\ref{fLyap}A shows the evolution of $\lmT$ during numerical simulations with different values of the passive forgetting rate $\lambda$. Its initial value ($L_1^{(1)} \approx 0.94$) is positive (we start our simulations with chaotic networks). As predicted by our theoretical approach (eq.\ref{eq:L1decay}), the Hebbian learning rule eq.(\ref{eq:LRule}) leads to a rapid decay of $\lmT$. The decay rate is indeed close to $\log(\| \WT \|)$ for intermediate learning epochs, in agreement with the upper bound of eq.(\ref{eq:L1decay}). Hence $\lmT$ quickly shifts to negative values, confirming the decrease of the dynamical complexity that could be inferred from visual inspection of fig.~\ref{fLyap}B.\\
One can also consider individual neuron activities. Figure~\ref{figu} shows the evolution of the local field $\u$ during learning. Clearly, the initial values are random, but the local field (thin gray line) shows a marked tendency to converge to the input pattern (thick dashed line) after as soon as $60$ learning epochs. At $T=180$, the convergence is almost complete. Hence this behavior once again conforms to the theoretical predictions eq.(\ref{eq:Uinf}), with $\XI \gg \Hinf$. In the results presented in this figure, we pursue the simulation up to $T=200$, at which point we remove the pattern from the network, i.e. we set $\xi_i=0, \, \forall i$ (fig.~\ref{fstruct}D). As a result, $\mathbf{u}$ looses its alignment from the pattern and presents a noisy aspect (note that each vector in the figure has been normalized to $[0, 1 ] $). This behavior is once again in agreement with the theoretical predictions of eq.(\ref{eq:Uinf}), which indicates that $\moyuinf = \Hinf$ upon pattern removal.\\
To conclude, we have shown here that Hebbian learning in our system leads to a decrease of the attractor complexity and entropy that can be induced by passive forgetting and/or an increased level of saturation of the neurons. This corresponds in details to the scenario predicted by our mathematical analysis.\\
\subsubsection{Functional consequences}
\label{sensitivity}
\begin{figure}
\centering
  \includegraphics[width=\textwidth]{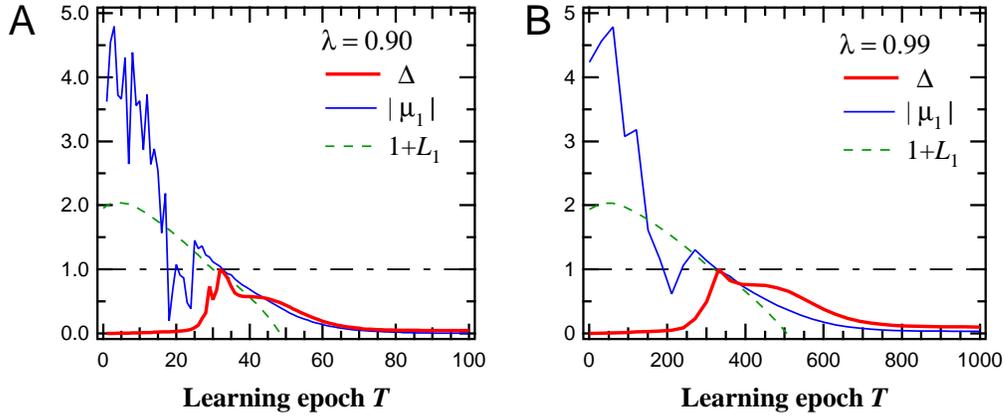}
\caption[long caption]{\label{Fsenspatt} Network sensitivity to the input pattern is maximal close to a bifurcation.
 The evolution  of the average value of the spectral radius
of $\DFx^{(T)}$ (thin full line) is plotted together with the sensitivity measure $\dtl$ (thick full line) for $\lambda=0.90$ (\emph{A}) or $0.99$ (\emph{B}). The panels also display the corresponding evolution of the largest Lyapunov exponent $L_1$, plotted as $1.0 + L_1$ for obvious comparison purpose (thin dashed line). The horizontal dashed-dotted lines locates $y=1$. The values of $\dtl$ are normalized to the $[0,1]$ range for comparison purposes.
 Each value is an average over 20 realizations (standard deviations are omitted for clarity). All other parameters were as in fig.~\ref{fstruct}}
\end{figure}

The former sections dealt with the effects of Hebbian learning on the structure and dynamics of the network. We now turn to its influence on the network function. The basic function of RRNNs is to learn a specific pattern $\XI$. In this framework, a pattern is learned when the complex (or chaotic) dynamics of the network settles onto a periodic oscillatory regime (a limit cycle) that is specific of the input pattern. This behavior emulates putative mechanisms of odor learning in rabbits that have been put forward by physiologists such as W. Freeman \cite{Freeman87,Freeman88}. An important functional aspect is that removal of the learned pattern after learning should lead to a significative change in the network dynamics. We now proceed to an analysis of this latter property.\\
The removal of $\XI$ is expected to change the attractor structure
and the average value of any observable $\phi$ (though with variable amplitude). Call $ \Delta^{(T)}\left[\phi\right]$ the variation of the (time) average value of $\phi$ induced by pattern removal. If the system is away from a bifurcation point, removal will result in a variation of
$\Delta^{(T)}\left[\phi\right]$ that remains proportional to $\XI$.\\
On the opposite, close to a bifurcation point this variation is typically not proportional
to $\XI$ and may lead to drastic changes in the dynamics.
From the analysis above, we therefore expect pattern removal
to have a maximal effect at ``the edge of chaos'', namely when the
(average) value of the spectral radius of $\DFx$ is close to $1$.
Call $\lambda_k$ and $\v_k$ the eigenvalues and eigenvectors of $\WT\Lambda(\uaT)$,
ordered such that $|\lambda_N| \leq |\lambda_{N-1}| \leq |\lambda_1|<1$. In the case where the dynamics has converged
to a stable fixed point $\uaT$ (namely, when
$\lmT <0$, see e.g. fig. \ref{fLyap}), our theoretical work predicted that:
\beq\label{DecSpecdu}
\Delta^{(T)}\left[\u\right] = -\sum_{k=1}^N \frac{\left(\v_k,\XI \right)}{1-\lambda_k}\v_k
\eeq
where $\left( \;,\; \right)$ denotes the inner product.  As a matter of fact, the RHS term diverges if $\lambda_1=1$ and if $\left(\v_1,\XI \right)\neq 0$. From this analysis, we therefore expect pattern removal
to have a maximal effect at ``the edge of chaos'', namely when the
 value of the spectral radius of $\DFx$ is close to $1$.\\
To study the effects of pattern removal in our model, we monitored the quantity:
\begin{equation}\label{eq:Delta}
\dtl=\frac{1}{N} \sqrt{
\sum_{i=1}^N
\left(
\lb \Lambda_{ii}(\u)\rbT
 - \lb\Lambda_{ii}(\u')\rbT
 \right)^2
}
\end{equation}
that measures how neuron excitability is modified when the pattern is removed.
The evolution of $\dtl$ during learning with rule eq.(\ref{eq:LRule}) is shown on fig.~\ref{Fsenspatt} (thick full lines)
for two values of the passive forgetting rate $\lambda$. $\dtl$ is found to increase to a plateau, and
vanishes afterwards. Interestingly, comparison with the decay of the leading eigenvalue of the Jacobian matrix, $\mu_1$ (thin full lines) shows that the maximal values of $\dtl$ are obtained when $|\mu_1|$ is close to $1$ and the largest Lyapunov exponent $L_1$ close to $0$.\\
Hence, these numerical simulations are in agreement with the theoretical predictions that \emph{Hebbian learning drives the global dynamics through a bifurcation, in the neighborhood of which sensitivity to the input
pattern is maximal.} Note that this property is obtained at the frontier where the chaotic strange attractor begins to destabilize ($|\mu_1|=1$), hence at the so-called ``edge of chaos''. This particular dynamical regime, at the frontier between order (periodical or fixed point regimes) and disorder (chaos), has already be reported to be particularly suitable for recurrent neural networks, especially when computational power is considered~\cite{Soula2005,Bertschinger04}. The present results show that it is the optimal regime for the sensibility to the input pattern in our model. Whether this also implies improved or optimal computational performance remains however to be tested and will be the subject of future works.\\
It must finally be noticed that our theory predicts that pattern sensitivity should be maximal when $|\mu_1|$ is close to one. But several aspects of our simulation results are not accounted for by this theory. For instance, fig.~\ref{Fsenspatt} shows that $|\mu_1|$ approaches $1$ at several learning epochs. This is related to the ``Arnold tongue'' structure of the route to chaos. However, pattern sensibility is maximal only for the last episode, and almost zero for the former ones. This behavior is still unclear and will be the subject of future works.\\
\section{Conclusion and future works}
To conclude, we have shown in this work that Hebbian learning eq.(\ref{eq:LRule}) has important effects on the dynamics and structure of a sparse two-populations RRNN. The forgetting part of the learning rule contracts the norm of the weight matrix. This effect, together with an increase in the average saturation level of the neurons, yields a rapid decay of the dynamics complexity and entropy. In other words, the network forgets its initial synaptic structure and is rewired by Hebbian learning into a new synaptic structure that emerges with learning and that depends on \textit{the whole history of the neuron dynamics}. We have shown that the strongest synapses organize within this emerging structure as a small-world connectivity. The second effect of the decrease of the weight matrix and of the increased neuron saturation consists in a rapid contraction of the spectral radius of the Jacobian matrix. This leads the system to the edge of chaos, where sensitivity to the input pattern is maximal. This scenario is remarkably predicted by the theoretical arguments we developed in~\cite{Siri07}.\\
In the presented simulations, most of the effects are mediated by the passive forgetting term. We believe that this term is not unrealistic from a biological point of view. Indeed, synaptic plasticity at the single synapse level is not permanent and some studied reported durations of 20 mn~\cite{Volianskis03} or even 20 sec~\cite{Brager03}. This would be accounted for in our model by $\lambda \ll 1$.\\
Nevertheless, most studies about long-term plasticity have evidenced longer cellular memory time constants, ranging from hours to days~\cite{Heynen00,Racine83,Doyere96}, which would correspond in our model to higher $\lambda$ values. Note however that according to our mathematical analysis, most of the effects reported here are expected to occur even without passive forgetting (i.e. with $\lambda = 1$), provided the learning rule increases the average saturation of the neurons. In previous studies, we have considered such Hebbian learning rules devoid of passive forgetting but provoking increasing average saturation levels of the neurons. Numerical simulations have clearly evidenced a reduction of the attractor complexity during learning, in agreement with this suggestion~\cite{Berry_AB06,Siri_ICCS06}.\\
Future works will focus on the study of more detailed biological learning rules (heterosynaptic LTD, synaptic rescaling). We will also consider activity-dependent synaptic turnover (pruning/sprouting). Indeed albeit an overlooked phenomena for several decades, synaptic (or at least dendritic) turnover is now recognized as an important part of cortical networks, even in the adult (see e.g.~\cite{Holtmaat05}). Finally, one important problem with the application of RRNNs as artificial neural networks, is that it is very difficult to determine when to stop the learning process. Our results show that the effect of an input pattern is maximal at those learning epochs when the system is close to a bifurcation, but much more modest for shorter \textit{and longer} learning times. One interesting development would thus consist in trying to find learning rules or settings that would guaranty that the system remains close to the edge of chaos, even at long learning times. As an attractive possibility, the plasticity of intrinsic properties~\cite{Daoudal03} could allow the network to stabilize its activity in this region.

\section*{Acknowledgments}
This work was supported by a grant of the French National Research Agency, project JC05\_63935 ``ASTICO''.

\bibliographystyle{abbrv}

\bibliography{biblio}

\newpage

\newpage

\newpage

\newpage

\newpage

\newpage

\newpage

\end{document}